\documentclass[aps,showpacs,nofootinbib,superscriptaddress]{revtex4}
\usepackage[dvips]{graphicx}
\usepackage{dcolumn}
\usepackage{multirow}

\begin{document}

\title{ Test of the heavy quark-light diquark approximation for baryons with a heavy quark} \author{ E. Hern\'andez}
\affiliation{Grupo de F\'\i sica Nuclear, Departamento de F\'\i sica
Fundamental e IUFFyM, Universidad de Salamanca, E-37008 Salamanca,
Spain.}  \author{J. Nieves} \affiliation{Departamento de F\'{\i}sica
At\'omica, Molecular y Nuclear, Universidad de Granada, E-18071
Granada, Spain.} \author{ J. M. Verde-Velasco} \affiliation{Grupo de
F\'\i sica Nuclear, Departamento de F\'\i sica Fundamental e IUFFyM,
Universidad de Salamanca, E-37008 Salamanca, Spain.}
\begin{abstract} 
  \rule{0ex}{3ex} We check a commonly used approximation in which a
  baryon with a heavy quark is described as a heavy quark-light
  diquark system.  The heavy quark influences the diquark internal
  motion reducing the average distance between the two light quarks.
  Besides, we show how the average distance between the heavy quark and any of
  the light quarks, and that between the heavy quark and the center of
  mass of the light diquark, are smaller than the distance between the
  two light quarks, which seems to contradict the heavy quark-light
  diquark picture. This latter result is in agreement with
  expectations from QCD sum rules and lattice QCD calculations. Our
  results also show that the diquark approximations produces larger
  masses than the ones obtained in a full calculation.

\end{abstract}

\maketitle

%
%
%
%
%
%
%
%
%

\section{Introduction}
 Heavy quark symmetry
\cite{nussinov87,shifman87,politzer88,isgur89,neubert94,korner94}
(HQS) predicts that in baryons with a heavy quark, and up to
corrections in the inverse of the heavy quark mass, the light degrees
of freedom quantum numbers are well defined, in particular the total
spin of the light degrees of freedom is well defined. This prediction
has been taken in different calculations as the basis for treating the
light quark subsystem as a diquark, and the baryon as a heavy
quark-light diquark (HQLD)
system~\cite{efimov91,guo93,guo96,ebert96,konig97,kaur99,guo01,ebert05,ebert06,guo07}.
This HQS prediction  does not imply though
 that the orbital motion of
the two light quarks is not affected by the presence of the heavy
quark as it seems to be implicit in the HQLD
approximation~\footnote{Note however that although in the HQLD approximation
the light diquark internal structure is not affected by the heavy
quark, this structure is commonly taken into account to build up the
heavy quark-light diquark interaction.}. 
Very recently the  diquark structure of heavy baryons have been analyzed in $\Lambda_c$
 production in heavy ion collisions~\cite{yasui08} where its enhanced yield is
 seen as a signal for the existence of light diquark correlations both in the quark gluon plasma and the heavy baryon.

In Ref.~\cite{cardarelli98}, using a light-front constituent quark model
and a Gaussian ansatz for the wave function, the authors studied the dependence of
the Isgur-Wise function~\cite{isgur89} on the baryon structure. They
found very different behaviors for a diquark-like configuration (the
heavy quark is far from the center of mass of the light quarks) or a
collinear-type configuration (the heavy quark is close to the center
of mass of the light quarks).  Comparison of the results with QCD sum
rules~\cite{grozin92} and lattice QCD calculations~\cite{ukqcd98}
suggested a clear dominance of the collinear-type configurations. 
This result seems to go against the HQLD approximation.

Here we plan to check the validity of the HQLD approximation, that we formulate in next section, by
 looking at heavy baryons masses and quark distributions inside 
 baryons composed of a heavy quark ($b$ or $c$) and two light
 quarks.  We shall compare the predictions
  obtained within that approximation with the ones obtained in
 a full calculation where the effect of the heavy quark on the light diquark is not neglected.
 For that purpose we shall use the nonrelativistic quark model
 and the full wave functions described in Ref.~\cite{albertus04}. In that
 reference we took advantage of HQS constraints on the spin of the
 light degrees of freedom to solve the full nonrelativistic three-body
 problem by means of a simple variational ansatz. The scheme of
 Ref.~\cite{albertus04} for the wave functions reproduced previous
 results for masses, charge radii\dots, obtained in
 Ref.~\cite{silvestre96} by solving more involved Faddeev
 equations. The baryons included in that and the present
 study appear in Table~\ref{tab:summ}.  We restrict ourselves to ground-state heavy baryons with total spin
 $J=1/2,\,3/2$ for which we could assume a zero total orbital angular
 momentum ($L=0$). 
\begin{table}[h!!!]
\begin{tabular}{cccccc||cccccc}\hline
Baryon &~~~~$S$~~~~&~~$J^P$~~&~~$I$~~&~~$S_{l}^\pi$~~~~& 
Quark content &~~Baryon~~ &~~$S$~~~~&~~$J^P$~~&~~$I$~~&~~$S_{l}^\pi$~~~~& 
Quark content
\\\hline
$\Lambda_c$& 0 &$\frac12^+$& 0 &$0^+$&$udc$&$\Lambda_b$& 0 &$\frac12^+$& 0 &$0^+$&$udb$\\
$\Sigma_c$ & 0 &$\frac12^+$& 1 &$1^+$&$llc$&$\Sigma_b$ & 0 &$\frac12^+$& 1 &$1^+$&$llb$\\
$\Sigma^*_c$ & 0 &$\frac32^+$& 1 &$1^+$&$llc$&$\Sigma^*_b$ & 0 &$\frac32^+$& 1 &$1^+$&$llb$\\
$\Xi_c$ & $-$1 &$\frac12^+$&$\frac12$&$0^+$&$lsc$ &$\Xi_b$ & $-$1 &$\frac12^+$&$\frac12$&$0^+$&$lsb$\\
$\Xi'_c$ & $-$1 &$\frac12^+$&$\frac12$&$1^+$&$lsc$&$\Xi'_b$ & $-$1 &$\frac12^+$&$\frac12$&$1^+$&$lsb$\\
$\Xi^*_c$ &$-$1&$\frac32^+$&$\frac12$&$1^+$&$lsc$&$\Xi^*_b$ &$-$1&$\frac32^+$&$\frac12$&$1^+$&$lsb$\\
$\Omega_c$ &$-$2 &$\frac12^+$& 0 &$1^+$&$ssc$&$\Omega_b$ &$-$2 &$\frac12^+$& 0 &$1^+$&$ssb$\\
$\Omega^*_c$ &$-$2 &$\frac32^+$& 0 &$1^+$&$ssc$&$\Omega^*_b$ &$-$2 &$\frac32^+$& 0 &$1^+$&$ssb$\\\hline
\end{tabular}
\caption{Summary of the quantum numbers of ground-state heavy baryons containing a single heavy quark. $I$, and
$S_{l}^\pi$ are the isospin, and the spin parity of the light
degrees of freedom and $S$, $J^P$ are the strangeness and the spin parity
of the baryon. We also give the quark content where $l$ denotes a light quark of flavor $u$ or
$d$.}
\label{tab:summ}
\end{table}
%
%
%
%
%
%
%
%
%
\section{heavy quark-light diquark approach to a heavy baryon}
\label{sect:hqld}
\begin{figure}[t]
\centerline{\resizebox{10.cm}{!}{\includegraphics{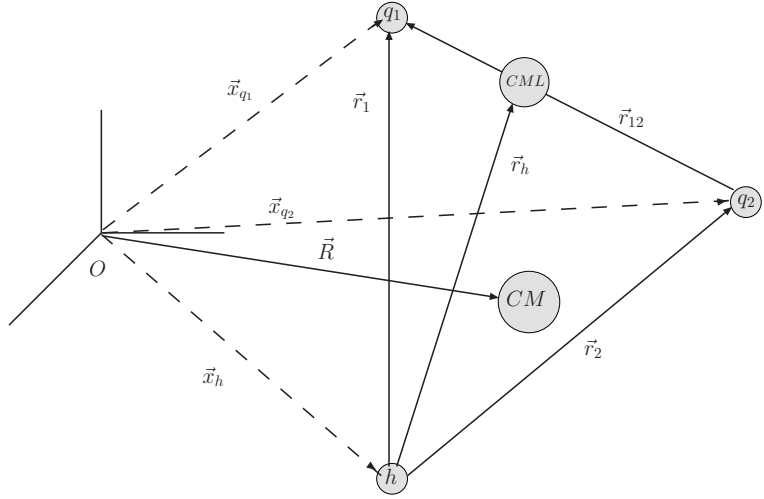}}}
\caption{\footnotesize Definition of different coordinates used
through this work. $CM$ and $CML$ stand for the baryon center of mass
and the light quark subsystem center of mass
respectively.}\label{fig:coor}
\end{figure}
The set of coordinates more adequate for
 a heavy quark-light diquark description are the Jacobi coordinates (See Fig.~\ref{fig:coor})

\begin{eqnarray}
\vec R &=&\frac{m_{q_1}\vec x_{q_1}+m_{q_2}\vec x_{q_2}+m_{h}\vec x_{h}}
{m_{q_1}+m_{q_2}+m_{h}}\nonumber\\
\vec r_{12}&=&\vec x_{q_1}-\vec x_{q_2}\nonumber\\
\vec r_h&=&\frac{m_{q_1}\vec x_{q_1}+m_{q_2}\vec x_{q_2}}{m_{q_1}+m_{q_2}}-\vec
x_h
\end{eqnarray} 
where $\vec x_{q_1},\, \vec x_{q_2}$ and $\vec x_h$ represent the
positions, with respect to a certain reference frame, of the two light
quarks and heavy quark respectively, and similarly $m_{q_1},\,
m_{q_2}$ and $m_h$ are their masses. The Jacobian coordinates are the
center of mass position $\vec R$, the relative position between the
two light quarks $\vec r_{12}$, and the relative position between the
two light quark center of mass and the heavy quark $\vec r_h$.
 
 In terms of these coordinates the three-body Hamiltonian can be written as
\begin{eqnarray}
H&=&
-\frac{\stackrel{\rightarrow}{\nabla}
\stackrel{}{^2}_{\hspace{-.1cm}\vec{R}}}{2 \overline M} +
H^{\rm int} \ \ ;\ \ \overline M=m_{q_1}+m_{q_2}+m_h\nonumber\\
 H^{\rm
int}&=&\overline M+H_{q_1q_2}+H_{hq_1q_2}
\end{eqnarray}
where $-\frac{\stackrel{\rightarrow}{\nabla}
\stackrel{}{^2}_{\hspace{-.1cm}\vec{R}}}{2 \overline M}$ accounts for
the total center of mass free motion. Besides $\overline M$, the
different terms in the internal Hamiltonian $H^{\rm int}$ are
\begin{eqnarray}
H_{q_1q_2}&=&-\frac{\stackrel{\rightarrow}{\nabla}_{12}^2}{2\mu_{q_1q_2}}
+V_{q_1q_2}(\vec r_{12},\, spin)\ \ \  \ ; \ \ \mu_{q_1q_2}=\frac{m_{q_1}m_{q_2}}{m_{q_1}+m_{q_2}}\nonumber\\
 H_{hq_1q_2}&=&-\frac12\left(\frac{1}{m_{q_1}+m_{q_2}}+\frac{1}{m_h}\right)
 \stackrel{\rightarrow}{\nabla}_{h}^2
 +V_{q_1h}(\vec r_h+\frac{m_{q_2}}{m_{q_1}+m_{q_2}}\vec r_{12},\,
spin)
+V_{q_2h}(\vec r_h-\frac{m_{q_1}}{m_{q_1}+m_{q_2}}\vec r_{12},\, spin)\nonumber\\
\end{eqnarray}
with $\stackrel{\rightarrow}{\nabla}_{12}=\partial /\partial_{\vec
r_{12}}$, $\stackrel{\rightarrow}{\nabla}_{h}=\partial /\partial_{\vec
r_h}$ and $V_{qq'}$ the interquark potential that depends on
relative distances and spins.  Defining now
\begin{eqnarray}
 H^0_{hq_1q_2}&=&-\frac12\left(\frac{1}{m_{q_1}+m_{q_2}}+\frac{1}{m_h}\right)
 \stackrel{\rightarrow}{\nabla}_{h}^2
+V_{q_1h}(\vec r_h,\,
spin)+V_{q_2h}(\vec r_h,\, spin)\nonumber\\
\end{eqnarray}
one could write
\begin{eqnarray}
H^{\mathrm{int}}=\overline M+H_{q_1q_2}+H^0_{hq_1q_2}+(H_{hq_1q_2}-H^0_{hq_1q_2})
\end{eqnarray}
$H_{q_1q_2}$ is the Hamiltonian for the relative motion of the two
 light quarks while $H^0_{hq_1q_2}$ is the Hamiltonian for the
 relative motion of the heavy quark with respect to a pointlike light
 diquark where the two light quarks are located in their center of
 mass.  Both Hamiltonians are coupled through the term
 $(H_{hq_1q_2}-H^0_{hq_1q_2})$. This latter term can not be neglected
 altogether as the light diquark is not pointlike.

Within the HQLD approximation one assumes that the light diquark 
internal structure is not disturbed by the presence of the heavy
quark. This means to neglect the influence of the term
$(H_{hq_1q_2}-H^0_{hq_1q_2})$  in the evaluation of the diquark
internal wave function, which therefore  will be determined by $H_{q_1q_2}$
alone. However, and since the diquark will have a finite size, the effect of
$(H_{hq_1q_2}-H^0_{hq_1q_2})$ has to be taken into account to obtain
the $r_h$ dependence of the baryon wave function and its mass.  Within this
approximation, we will take a baryon wave function  given by
\begin{equation}
\Psi^{B,\,HQLD}_{hq_1q_2}(r_{12},r_h)=\Phi_{q_1q_2}(r_{12})\cdot F_{hq_1q_2}(r_h)
\label{eq:wfhqld1}
\end{equation}
 where $\Phi_{q_1q_2}(r_{12})$  is the ground-state wave function for
 the Hamiltonian  $H_{q_1q_2}$ and the given spin configuration. We
 will determine  $F_{hq_1q_2}(r_h)$ variationally assuming an ansatz
 of the form
 \begin{equation}
F_{hq_1q_2}(r_h)=\Phi^0_{hq_1q_2}(r_h)\cdot N\bigg(1+\sum_{j=1}^2\,a_j\,e^{-b_j^2\,(r_h+c_j)^2}\bigg)
\label{eq:wfhqld2} 
\end{equation}
 with $\Phi^0_{hq_1q_2}(r_h)$ the ground-state wave function for
 $H^0_{hq_1q_2}$ for the given spin configuration\footnote{
 $\Phi^0_{hq_1q_2}(r_h)$ and $\Phi_{q_1q_2}(r_{12})$ can be easily
 obtained by solving the corresponding Schr\"odinger equations with a
 Numerov algorithm.}, and where $N$ is a normalization constant and
 $a_j,\,b_j,\,c_j;\ j=1,2$ are variational parameters that we
 determine by energy minimization. The variational parameters 
 are compiled in  the appendix.
%
%
%
%
\section{Comparison of the diquark approximation with the full calculation}
As stated in the introduction, our full calculation in
Ref.~\cite{albertus04} took advantage of HQS constraints on the total
spin of the light quarks to solve the full three-body problem by using
a variational ansatz. All the information on the wave functions can be
found there. In this section we shall compare results for masses and
quark distributions obtained with the full calculation and with the
HQLD approximation corresponding to
Eqs.~(\ref{eq:wfhqld1}-\ref{eq:wfhqld2}).  All the results that we
shall present have been obtained with the use of the AL1 interquark
interaction of Refs.~\cite{semay94,silvestre96}. This interaction
contains a confinement term plus Coulomb and hyperfine terms coming
from one gluon exchange. It was initially developed for mesons and for
its use in baryons we have applied the usual $V_{qq}=V_{q\bar q}/2$
prescription~\cite{silvestre96,badhuri81}.

\begin{table}[t]
\begin{tabular}{cccc||cccc}\hline
Baryon &\ \ Full calcul.~\cite{albertus04}\ \ \ \ & HQLD approx. & Exp.&\ \  Baryon &\ \ Full calcul.~\cite{albertus04}\ \ &\ \  
HQLD approx.& Exp.\ \ \\\hline
$\Lambda_c$& 2295& 2317&$2286.48\pm0.14$&$\Lambda_b$&5643&5663&$5624\pm9$ \\
$\Sigma_c$ &2469& 2521&$2453.6\pm0.5$&$\Sigma_b$ &5851&5897&$5812^\S\pm3$\\
$\Sigma^*_c$ & 2548&2579 &$2518\pm2$&$\Sigma^*_b$&5882&5919&$5833^\S\pm3$ \\
$\Xi_c$ & 2474&2501&$2469.5\pm0.6$ &$\Xi_b$ &5808&5837& $5760^\ddag\pm70$ \\
$\Xi'_c$ & 2578&2629&$2577\pm4$&$\Xi'_b$ & 5946&5993&$5900^\ddag\pm70$\\
$\Xi^*_c$ &2655&2686&$2646\pm1.4$&$\Xi^*_b$ &5975&6015&$5900^\ddag\pm70$\\
$\Omega_c$ &2681&2727&$2697.5\pm2.6$&$\Omega_b$&6033&6081&$5990^\ddag\pm70$\\
$\Omega^*_c$&2755&2783&$2768.3^\dag\pm3.2$&$\Omega^*_b$&6063&6104&$6000^\ddag\pm70$  \\\hline
\end{tabular}
\caption{Masses in MeV obtained with our full calculation in
Ref.~\cite{albertus04} and with the HQLD approximation (See text for
details).  In all cases we use the AL1 interquark potential of
Refs.~\cite{semay94,silvestre96}. We also show experimental masses
(isospin average) and lattice estimates when the former are not known.
Experimental masses have been taken from Refs.~\cite{pdg06},
\cite{babar06} ($\dag$) and \cite{cdf07}($\S$).  Lattice estimates
($\ddag$) have been taken from Ref.~\cite{ukqcd96}. Note in
Ref.~\cite{babar06} what it is actually measured is the mass
difference $M_{\Omega^*_c}-M_{\Omega_c}$.}
\label{tab:masses}
\end{table}
In Table~\ref{tab:masses} we compare the masses obtained as explained
 above.  The masses of the full calculation are smaller in all cases,
 and thus better from a theoretical point of view\footnote{For a given
 Hamiltonian a variational wave function gives an upper limit to the
 ground-state mass}. They also compare better with
 experiment~\cite{pdg06,babar06,cdf07} and lattice
 estimates~\cite{ukqcd96}.  The results show that a full calculation
 makes the whole system more bound producing smaller masses.

The differences we have seen in the masses are a reflection of
differences in the wave functions.  We now make direct comparisons
between the wave functions in the full calculation and in the HQLD
approximation.  We start by looking at the projection ${\cal P}$ of
our full variational wave functions $\Psi^B_{hq_1q_2}(r_1,r_2,r_{12})$
obtained in Ref.~\cite{albertus04} onto
$\Psi^{B,\,HQLD}_{hq_1q_2}(r_{12},r_h)$.  Those projections are given
by\footnote{Note in Ref.~\cite{albertus04} the variational wave
functions were obtained using a different set of coordinates which are
$\vec r_{12}$, $\vec r_1=\vec
r_h+\frac{m_{q_2}}{m_{q_1}+m_{q_2}}\,\vec r_{12}$ and $\vec r_2=\vec
r_h-\frac{m_{q_1}}{m_{q_1}+m_{q_2}}\,\vec r_{12}$.\ \ $\vec r_1$ and
$\vec r_2$ are the relative coordinates of the two light quarks with
respect to the heavy quark. Besides note that
$d^3r_1d^3r_2=d^3r_{12}d^3r_h$}
\begin{equation}
{\cal P}=\int d^3r_1 \int d^3r_2 \left(\Psi^B_{hq_1q_2}(r_1,r_2,r_{12})\right)^*
\Psi^{B,\,HQLD}_{hq_1q_2}(r_{12},r_h)
\label{eq:project}
\end{equation}
and the $|{\cal P}|^2$ values give an idea of how much of the ``true''
wave function is given by the wave function of the HQLD
approximation. The values for $|{\cal P}|^2$ appear in
Table~\ref{tab:project}.
\begin{table}[h!!!]
\centering
\begin{tabular}{lcccccccc}
&\ \ $\Lambda_c$\ \ &\ \ $\Sigma_{c}$\ \ &\ \ $\Sigma^*_{c}$\ \ &\ \ $\Xi_c$\ \ &\ \ $\Xi'_c$\ \ &\ \ $\Xi^*_c$\ \ &\ \ $\Omega_c$\ \  &\ \ $\Omega^*_c$\\
\hline\ \ 
$|{\cal P}|^2$ &0.971&0.943&0.957&0.949&0.926& 0.932&0.935&0.961\\\hline
\\\\
&$\Lambda_b$&$\Sigma_{b}$&$\Sigma^*_{b}$&$\Xi_b$&$\Xi'_b$&$\Xi^*_b$&$\Omega_b$ &$\Omega^*_b$\\
\hline
$|{\cal P}|^2$ &0.949&0.946&0.951&0.924&0.921&0.922&0.935&0.946\\\hline
\end{tabular}
\caption{Absolute value square of the ${\cal P}$ projection coefficient  defined in
Eq.~(\ref{eq:project})}
\label{tab:project}
\end{table}%

The values are generally higher for $c$-baryons than for the
corresponding $b$-baryons. For a given isospin and strangeness the
larger values occur for baryons with $S_l=0$, and in the cases where
$S_l=1$ the projection is maximum for states with total angular
momentum $3/2$. To get more insight into what could be left in the
3-8\% discrepancy that one observes we have evaluated different quark
distributions.

It is very interesting to compare the probability $P_{q_1q_2}$ for the
two light quarks to be found at a relative distance $r$. This
probability is evaluated as
\begin{equation}
P_{q_1q_2}\,(r) =\int d^3r_1\int d^3r_2\ \delta(r_{12}-r)
\left|\Psi^B_{hq_1q_2}(r_1,r_2,r_{12})
\right|^2
\label{eq:pq1q2}
\end{equation}
in the full calculation, and more simply as 
\begin{equation}
\left.P_{q_1q_2}\,(r)\right|_{HQLD} =4\pi\,r^2\,\left|\Phi_{q_1q_2}(r)\right|^2
\label{eq:pq1q2dq}
\end{equation} 
 in the HQLD approximation case\footnote{Note in the HQLD
 approximation $P_{q_1q_2}$ is totally independent of
 $F_{hq_1q_2}(r_h)$.}.

\begin{figure}[t]
\center
\resizebox{12.cm}{17cm}{\includegraphics{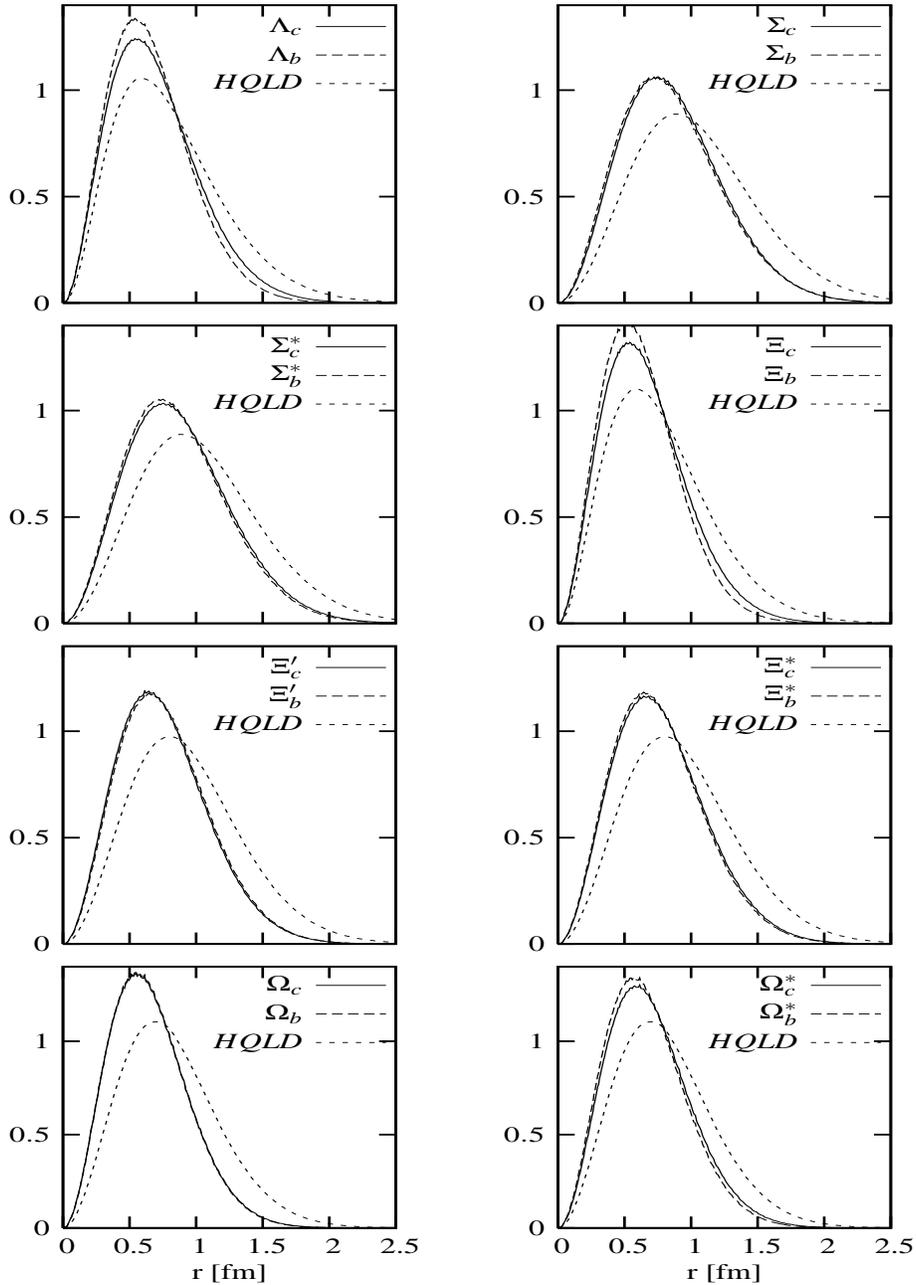}}
\caption{$P_{q_1q_2}\,(r)$ defined in
Eqs.~(\ref{eq:pq1q2}-\ref{eq:pq1q2dq}) evaluated using our full
calculation (solid lines for $c$-baryons and long-dashed lines for
$b$-baryons) or with the HQLD approximation (short-dashed lines).}
\label{fig:p12}
\end{figure}

The results of $P_{q_1q_2}(r)$ for different $b$- and $c$-
ground-state heavy baryons appear in Fig.~\ref{fig:p12}. In the HQLD
approximation the results do not depend on the heavy quark mass, only
on the quark content of the diquark and on the $S_l$ value.  This
feature is shared by the full calculation where one sees little
dependence on the heavy quark mass. On the other hand we see in the
full calculation how the presence of the heavy quark affects the
diquark internal structure decreasing the relative distance between
the two light quarks making the whole system more bound.  This is the
expected behavior from the comparison of the masses.

\begin{figure}[t]
\center
\resizebox{12.cm}{17cm}{\includegraphics{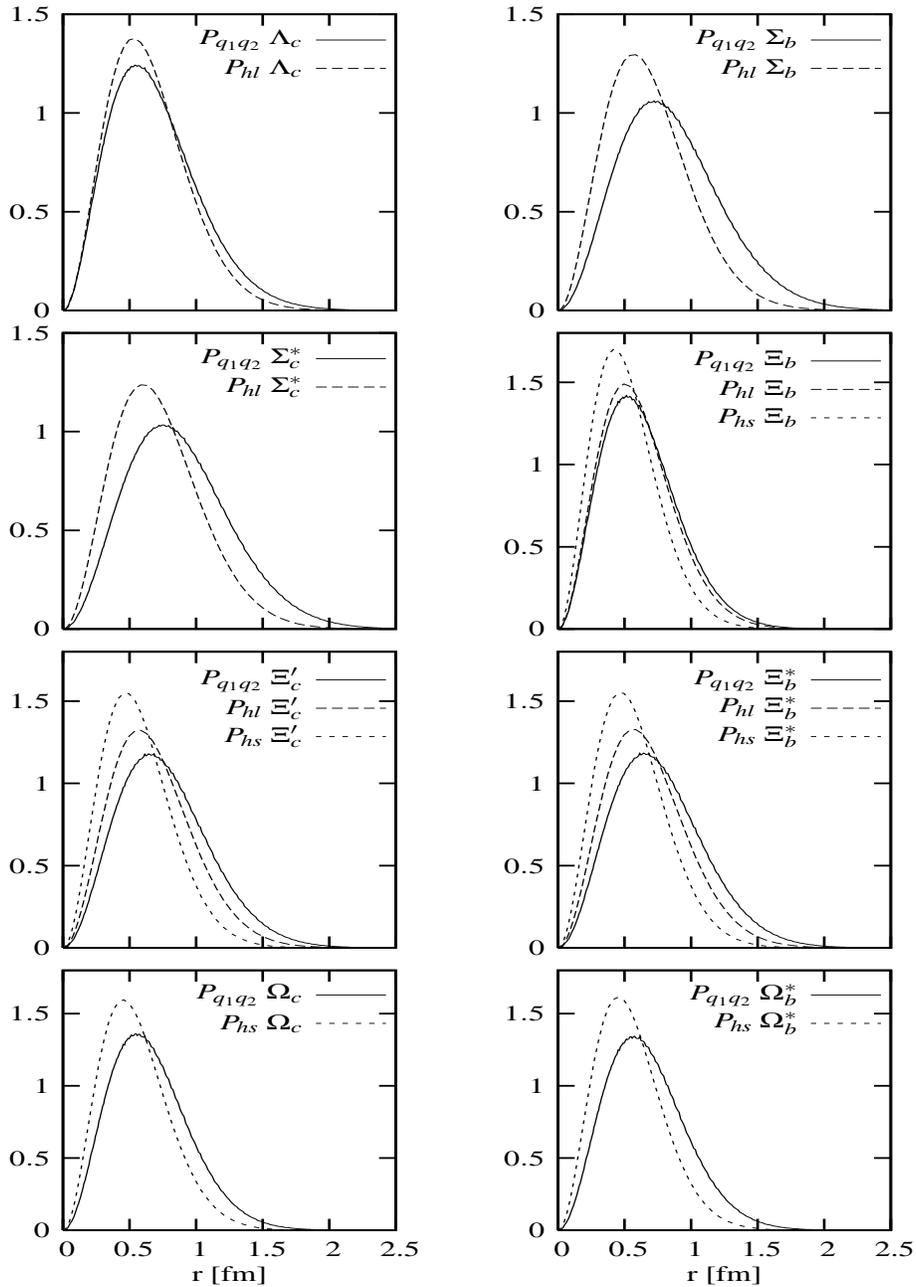}}
\caption{$P_{hq_j}\,(r)$ (long-dashed lines for $q_j=l$ and short-dashed lines for $q_j=s$) and $P_{q_1q_2}\,(r)$ (solid lines) 
evaluated using  our full calculation.}
\label{fig:phqj}
\end{figure}

Another piece of information is provided by the full calculation
probability $P_{hq_j}(r)$ to find the heavy quark at a certain
distance $r$ of a light quark
\begin{eqnarray}
&&P_{hq_j}\,(r) =\int d^3r_1\int d^3r_2\ \delta(r_j-r)\
\left|\Psi^B_{hq_1q_2}(r_1,r_2,r_{12})
\right|^2
\label{eq:phqj}
\end{eqnarray}
The results are shown in Fig.~\ref{fig:phqj} where for comparison we
also show the corresponding $P_{q_1q_2}(r)$ distribution. From the
figure one sees the heavy quark is closer to any of the two light
quarks than the latter two among themselves.

\begin{figure}[t]
\center
\resizebox{12.cm}{17cm}{\includegraphics{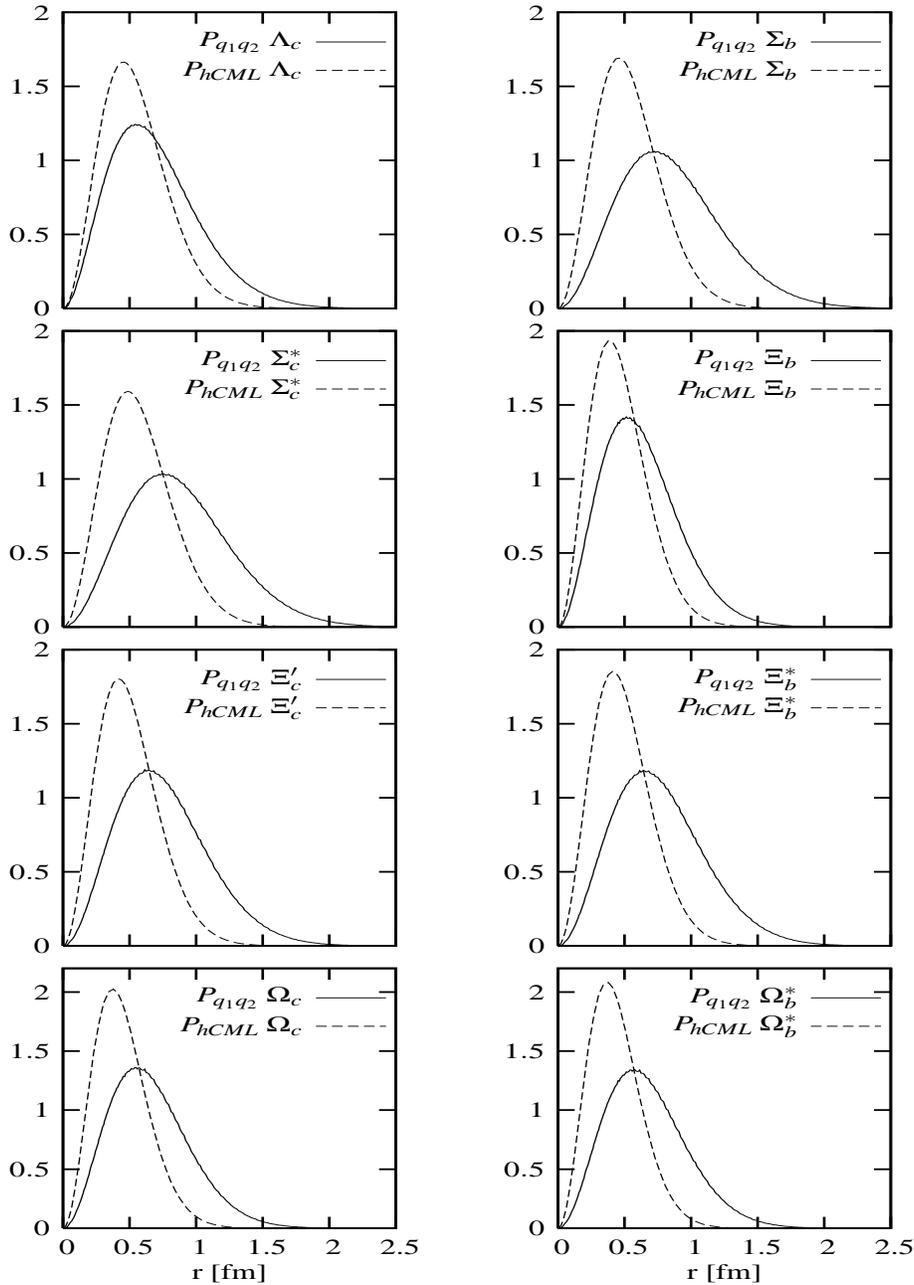}}
\caption{$P_{hCML}\,(r)$ (dashed lines) and $P_{q_1q_2}\,(r)$ (solid
lines) evaluated using our full calculation .}
\label{fig:phcml}
\end{figure}

Finally, we have also evaluated the probability distribution
$P_{hCML}(r)$ for the heavy quark to be at a certain distance $r$ of
the center of mass of the two light quarks $CML$. Again, this is
simply given by
\begin{eqnarray}
&&P_{hCML}\,(r) =\int d^3r_1\int d^3r_2\ \delta(r_h-r)\
\left|\Psi^B_{hq_1q_2}(r_1,r_2,r_{12})\right|^2
\label{eq:phcml}
\end{eqnarray}
and the results are shown in Fig.~\ref{fig:phcml}, where we also show
the $P_{q_1q_2}(r)$ distributions. What one sees is that the average
distance of the heavy quark to the center of mass of the light degrees
of freedom is smaller than the average distance between the two light
quarks.

\begin{figure}[h!!]
\center
\resizebox{5.cm}{!}{\includegraphics{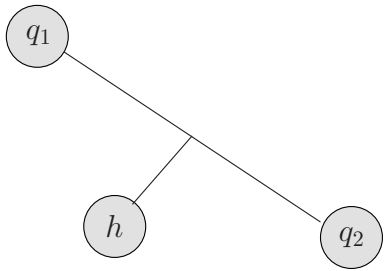}}
\caption{Emerging schematic picture of a baryon with a heavy quark. This is in agreement wit the findings of Ref.~\cite{cardarelli98} (See text for details).}
\label{fig:colineal}\vspace{.5cm}
\end{figure}

\begin{figure}[h!!]
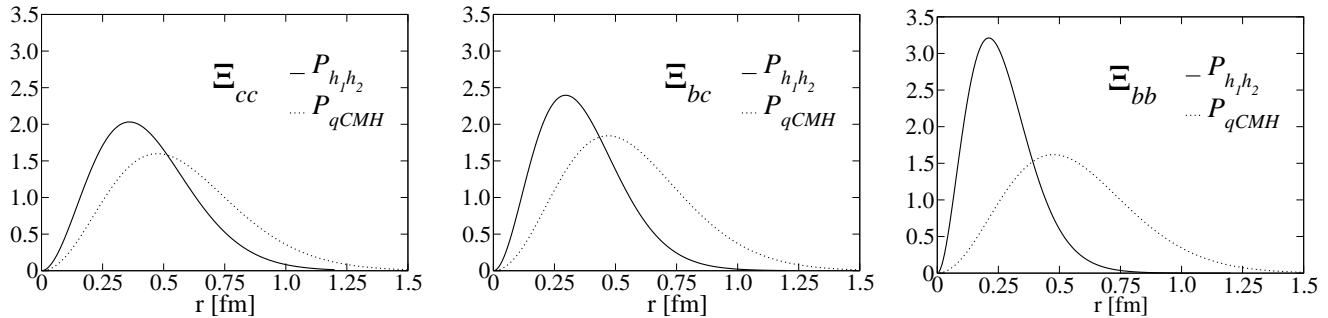

\resizebox{5.5cm}{!}{\includegraphics{xi_cc.eps}}\hspace{.4cm} 
\resizebox{5.5cm}{!}{\includegraphics{xi_bc.eps}}\hspace{.4cm}\resizebox{5.5cm}{!}{\includegraphics{xi_bb.eps}}
\caption{Probability distribution $P_{h_1h_2}(r)$ (solid lines) for two heavy quarks
to be found at a certain distance, and probability distribution
$P_{qCMH}(r)$ (dotted lines) for the light quark to be found at a distance of the two
heavy quark center of mass $CMH$ evaluated for doubly heavy $\Xi$
baryons. We have used our full wave functions from Ref.
\cite{albertus07}.}
\label{fig:dp}
\end{figure}

The picture that emerges from this analysis is the one depicted in
Fig.\ref{fig:colineal}, where the heavy quark is too close to the
center of mass of the light degrees of freedom for the HQLD
approximation to be fully valid. This result confirms the findings of
Ref.~\cite{cardarelli98}. There the comparison of the Isgur-Wise
functions, obtained for different heavy baryon configurations, with
the results of QCD sum rules~\cite{grozin92} and lattice
calculations~\cite{ukqcd98} showed a dominance of this collinear-type
configuration. By contrast, for doubly heavy baryons the light
quark-heavy diquark picture is clearly favored. We illustrate this
point in Fig.~\ref{fig:dp} for the case of doubly heavy $\Xi$
baryons. There we show the probability distribution $P_{h_1h_2}(r)$
for the two heavy quarks to be at a certain distance, and the
probability distribution $P_{qCMH}(r)$ for the light quark to be found
at a certain distance of the two heavy quark center of mass $CMH$. For
the evaluation we have used our full wave functions obtained in
Ref.~\cite{albertus07}. We see how as the heavy quark masses increase
the maximum of $P_{h_1h_2}(r)$ moves to lower distances while for
$P_{qCMH}(r)$ the maximum does not change.

\section{Concluding remarks}
We have checked the HQLD approximation by looking at its effects on
masses and quark distributions inside the baryon. In that
approximation the baryon is described as a bound state of a heavy
quark and a light diquark which internal structure is not affected by
the presence of the heavy quark. The approximation seems to work
reasonably well at the level of total masses, although a full
calculation produces smaller mass values. On the other hand our
results show that the presence of the heavy quark affects notably the
relative motion of the light degrees of freedom reducing the average
distance between the two light quarks. Besides one sees that the heavy
quark is closer to the light quarks than the latter among themselves,
and that its average distance to the center of mass of the light
quarks is also smaller than the size of the diquark. All this
information seems to contradict the heavy quark-light diquark picture.
Our study confirms previous analysis on the structure of heavy baryons
done in Ref.~\cite{cardarelli98}. Similar results concerning the quark
distributions are obtained in the relativistic quark model of Ebert
{\it et al.}~\cite{ebertpc}.  The use of a full calculation seems to
be preferable.

%
%
%
%
%
%
%
%

\begin{acknowledgments}
We thank Prof. D. Ebert for suggesting us to do this analysis.
               This research was supported by DGI and FEDER funds, under contracts
FIS2005-00810,  FIS2006-03438, FPA2007-65748, and the Spanish Consolider-Ingenio 2010
Programme CPAN (CSD2007-00042), by Junta de
Andaluc\'\i a and Junta de Castilla y Le\'on under contracts FQM0225
and SA016A07, and it is part of the EU integrated infrastructure
initiative Hadron Physics Project under contract number
RII3-CT-2004-506078.  
\end{acknowledgments}

%
%
%
%
%
%
%
%
%
\appendix
\section{}
In Table~\ref{tab:param} we give the values for the $a_j,\,b_j,\,c_j;\ j=1,2$ parameters of Eq.~(\ref{eq:wfhqld2}) that minimize the masses
in the HQLD approximation.
\begin{table}[h!!]
\begin{tabular}{lcccccc||lcccccc}\hline
&\ $a_1$\  &\  $b_1\, [fm^{-1}]$\  &\ $c_1\, [fm]$\  &\  $a_2$\  &\ $b_2, [fm^{-1}]$\  &\  $c_2\, [fm]$\ 
& &\  $a_1$\  &\  $b_1, [fm^{-1}]$\  &\  $c_1\, [fm]$\  &\  $a_2$\  &\  $b_2, [fm^{-1}]$\  &\  $c_2\, [fm]$\ 
\\\hline
$\Lambda_c$\hspace{.5cm}& $-0.236$ & 0.563 & 0.253 &$-0.499$&  0.537&  0.505&\ \ $\Lambda_b$\hspace{.5cm}&$-0.493$&  0.524&  0.921& $-0.427$&  0.524& 
0.998\\
$\Sigma_c$ &$-0.475$  & 0.330  & 0.551 & $-0.484$  & 0.330  & 0.553&\ \ $\Sigma_b$&$-0.376$ & 0.211&  0.851& $-0.618$&  0.190 & 1.014 \\ 
$\Sigma^*_c$ &$-0.250$ & 0.560 & 0.253 &$-0.502$ & 0.535 & 0.505&\ \ $\Sigma^*_b$ &$-0.384$  & 0.397 &  0.822 & $-0.560$  & 0.380  &
0.980\\
$\Xi_c$ & $-0.589$ &  0.544 &  0.917&  $-0.330$&   0.555&   0.998&\ \ $\Xi_b$ &$-0.456$ & 0.507 & 0.797 &$-0.511$&  0.497 & 0.965  \\
$\Xi'_c$ &$-0.478$ & 0.344&  0.547 &$-0.486$ & 0.344 & 0.549&\ \ $\Xi'_b$ &$-0.477$&  0.392 & 0.521 &$-0.489$ & 0.391 & 0.533 \\
$\Xi^*_c$ &$-0.569$&  0.499 & 0.928 &$-0.348$ & 0.507&  1.019&\ \ $\Xi^*_b$ &$-0.359$ & 0.421&  0.769 &$-0.634$ & 0.398 & 0.958\\
$\Omega_c$ &$-0.474$ & 0.396 & 0.516& $-0.489$ & 0.395 & 0.532&\ \ $\Omega_b$ &$-0.487$ & 0.408&  0.528 &$-0.491$ & 0.408 & 0.529\\
$\Omega^*_c$ &$-0.435$ & 0.583 & 0.856& $-0.525$&  0.573 & 0.967&\ \ $\Omega^*_b$ &$-0.360$ & 0.419 & 0.697& $-0.652$ & 0.393 & 0.925\\\hline
\end{tabular}
\caption{}
\label{tab:param}
\end{table}%
%
%
%
%
%
%
%
%

\end{document}